\documentclass[usenatbib,onecolumn]{mn2e}

\usepackage{amsmath,amssymb}
\usepackage{graphicx}

\def\Ep{E^{\scriptscriptstyle(+)}}
\def\Em{E^{\scriptscriptstyle(-)}}
\def\Epm{E^{\scriptscriptstyle(\pm)}}
\def\Sp{S^{\scriptscriptstyle(+)}}
\def\Sm{S^{\scriptscriptstyle(-)}}
\def\Spm{S^{\scriptscriptstyle(\pm)}}
\def\Lp{L^{\scriptscriptstyle(+)}}
\def\Lm{L^{\scriptscriptstyle(-)}}
\def\Lpm{L^{\scriptscriptstyle(\pm)}}

\def\tcoh{\Delta\tau}
\def\dt{\Delta t}
\def\snr{{\rm SNR}}

\def\up#1{\raise.5ex\hbox{#1}}
\def\down#1{\lower.8ex\hbox{#1}}

\title{Intensity interferometry with more than two detectors?}

\author[Malvimat, Wucknitz \& Saha]
{Vinay Malvimat,$^1$ Olaf Wucknitz$^{2,3}$ and Prasenjit Saha$^4$ \\
$^1$ School of Physics, University of Hyderabad, Gachibowli,
  Hyderabad-500046, India\\
$^2$ Max-Planck-Institut f\"ur Radioastronomie, Auf dem H\"ugel 69,
53121 Bonn, Germany\\
$^3$ Argelander-Institut f\"ur Astronomie, Auf dem H\"ugel 71,
53121 Bonn, Germany\\
$^4$ Institute for Theoretical Physics, University of Zurich,
Winterthurerstr~190, 8057 Zurich, Switzerland \\
}
\date{Accepted 2013 October 8.  Received 2013 September 10; in original form 2013 April 11}

\begin{document}

\maketitle

\label{firstpage}

\begin{abstract}
The original intensity interferometers were instruments built in the
1950s and 60s by Hanbury Brown and collaborators, achieving
milli-arcsec resolutions in visible light without optical-quality
mirrors.  They exploited a then-novel physical effect, nowadays known
as HBT correlation after the experiments of Hanbury Brown and Twiss,
and considered fundamental in quantum optics.  Now a new generation of
intensity interferometers is being designed, raising the possibility
of measuring intensity correlations with three or more detectors.

Quantum-optics predicts two interesting features in many-detector HBT:
(i)~the signal contains spatial information about the source (such as
the bispectrum or closure phase) not present in standard HBT, and
(ii)~correlation increases combinatorially with the number of
detectors.  The signal to noise ratio (SNR) depends crucially on the
number of photons --- in practice always $\ll 1$ --- detected per
coherence time.  A simple SNR formula is derived for thermal sources,
indicating that three-detector HBT is feasible for bright stars.  The
many-detector enhancement of HBT would be much more difficult to
measure, but seems plausible for bright masers.
\end{abstract}


\section{Introduction}

If two detectors are counting photons from an ordinary incoherent
light source, there is a tendency for photons to arrive at both
detectors together.  This is known as photon bunching or HBT
correlation, after the pioneering work of Hanbury Brown and Twiss, and
is a consequence of the bosonic quantum statistics of photons.  In a
situation where coherent light would produce interference fringes, the
HBT correlation with incoherent light varies according to those
would-be but absent fringe patterns.  This allows a type of
interferometry with incoherent light and without optical-quality
mirrors, known as intensity interferometry.

HBT correlation appears naturally in classical wave optics, and was
actually first used to build a radio intensity interferometer, which
resolved an extra-galactic radio source for the first time
\citep{1952Natur.170.1061H}.  But when the same ideas were applied to
visible light, and the effect was measured, first in the lab
\citep{1956Natur.177.27H} and then with starlight
\citep{1958RSPSA.248..222B}, it became controversial, because it
implied that different photons could interfere, contrary to
conventional wisdom at the time.  The controversies were eventually
resolved with the development of a quantum-statistical theory for
incoherent light by \cite{PhysRevLett.10.277} and
\cite{PhysRev.131.2766} and the emergence of quantum optics.  In
quantum optics, ordinary light (or `chaotic light') behaves like a
random mixture of lasers, and the semi-classical picture of
interfering waves, the squared amplitudes of which determine the
emission rate of photo-electrons, turns out to be valid.\footnote{The
  semi-classical description is not valid in general.  A nice
  counter-example is provided by electrons, which show HBT {\em
    anti\/}-correlation \citep{2002Natur.418..392K} from fermion
  statistics.}  The general phenomenon of photon bunching or HBT
correlation then spread to different areas of physics in different
guises. For example, it is well known in the context of nuclear
collisions \citep{1998AcPPB..29.1839B}, and it may even be relevant to
animal vision \citep{PhysRevLett.109.113601}.

Meanwhile, Hanbury Brown and collaborators developed the Narrabri
Stellar Intensity Interferometer (NSII) which measured stellar
diameters down to milli-arcsecs \citep{1968ARA&A...6...13B}.  But
the photon detectors then available were only blue-sensitive, limiting
the instrument to hot stars.  So the NSII ran out of stars to observe
in a few years, after which it was dismantled and almost forgotten.
Decades later, with a new generation of detectors, intensity
interferometry has become somewhat topical in astronomy again.  A
number of proposals and experiments have appeared in recent years
\citep{2006MNRAS.368.1646O,2006MNRAS.368.1652O,2008MNRAS.389..364B,2008A&A...484..887J,2009A&A...507.1719F,2012SPIE.8445E..2LH}.
Particularly ambitious are plans to adapt Cherenkov telescopes into a
giant reincarnated NSII with resolution down to $<0.03\rm\,mas$
\citep{2012MNRAS.419..172N,2012NewAR..56..143D}.

Future intensity interferometers are likely to have many detectors,
not just two.  Clearly, detectors can work in simultaneous pairs.  The
prospect of measuring $N$-fold photon coincidences has also been
suggested \citep[e.g.,][]{2006MNRAS.368.1646O}.  The basic theory is
well established in the quantum-optics literature and three-detector
HBT has been measured in laboratory experiments
\citep{Sato:78,PhysRevA.81.043831}.  But are the results interesting for an
astronomical instrument?  This paper examines the question.

\section{Standard HBT}

A nice introduction to HBT from a modern quantum-optics perspective
appears in \cite{RevModPhys.78.1267}.  We quote some key results here.

The central concept is the field correlation functions.  Let $x_1$ and
$x_2$ denote the spacetime locations of two detectors.  The
first-order correlation\footnote{In a different terminology our
  `first-order correlation' would be called `two-point correlation',
  and our `second-order correlation' is a `four-point correlation'.}
is defined as the average
\begin{equation}
G(x_1,x_2) \equiv \big\langle \Em(x_1) \, \Ep(x_2) \big\rangle \,\down.
\end{equation}
Classically, $\Epm$ denote the positive and negative frequency parts
of the electric field at a detector.  In quantum statistics, the
fields become operators.  We recognize $G(x_1,x_2)$ as the unnormalized
correlation function or `visibility' in radio
interferometry, or the spatial Fourier transform of the source in the
sky. Note that $G(x_2,x_1)$ is the complex conjugate (Hermitian
conjugate in quantum statistics) of $G(x_1,x_2)$.  Thus $G(x_1,x_1)$,
which is the count rate at $x_1$, is automatically real and
non-negative.  For a laser source, $G(x_1,x_2)$ would be independent
of time, but for chaotic sources, the correlation falls away over a
coherence time
\begin{equation} \label{ctime}
\tcoh \approx 1/\Delta\nu \,.
\end{equation}
Now consider a form of second-order correlation
\begin{equation}
G^{(2)}(x_1,x_2,x_2,x_1) \equiv 
\big\langle \Em(x_1) \, \Em(x_2) \, \Ep(x_2) \, \Ep(x_1) \big\rangle
\end{equation}
which is the probability of coincident detection within $\tcoh$.
Chaotic sources have the property (cf.~Glauber's equation~37)
\begin{equation} \label{G37}
G^{(2)}(x_1,x_2,x_2,x_1) = G(x_1,x_1) \, G(x_2,x_2) + |G(x_1,x_2)|^2 \,.
\end{equation}
The first term corresponds to the
random coincidences that would also be expected for classical
particles, but the second term describes the non-classical HBT
correlations.

To estimate the signal to noise, suppose we have two detectors with
equal count rates $r$ (photons per time unit), parametrized as
\begin{equation}
r\,\tcoh = G(x_1,x_1) = G(x_2,x_2) \,,
\end{equation}
so that $G(x,x)$ provides the number of photons per coherence time.
Let us now count photons over some time $\dt$. The time resolution
$\dt$ of the detector, typically $\dt \gg \tcoh$, is often called the
reciprocal electrical bandwidth, because in the early experiments it
was set by amplifier properties.

The product of photon counts (corresponding to random coincidences)
will be close to $(r\dt)^2$, as given by the first term on the right
of equation~(\ref{G37}). The HBT signal given by the second term will
make a small but non-zero contribution. Over a coherence time it will
be $|G(x_1,x_2)|^2$.  Over the much longer interval $\dt$ there are,
so to speak, $\dt/\tcoh$ coherent slices, making the HBT signal
$r^2\tcoh\,\dt$.  Using Poisson noise for the photon numbers and thus
interpreting the square root of the total coincidence rate as the
noise, we get a signal to noise ratio
\begin{equation} \label{basic-snr}
\snr (\dt) \sim r\,\tcoh
\end{equation}
for full correlation.  This applies to a single counting time
$\dt$. Over many counting times, the \snr\ adds in quadrature so that
we get\footnote{It is understood that
  \snr\ must also include factors for throughput and detector
  efficiency \citep[cf.\ equation 14 in][]{1968ARA&A...6...13B}.  But
  equation \eqref{basic-snrT} is the essential \snr\ expression in
  standard HBT.}
\begin{equation} \label{basic-snrT}
\snr (T) \sim r\,\tcoh \sqrt\frac{T}{\dt}
\end{equation}
when integrating over a duration $T$.When increasing the coherence
time by using narrow-band filters, the photon rate $r$ will decrease
with $\Delta\nu\approx1/\tcoh$ so that the `spectral intensity'
$r\tcoh$ stays constant. This has the remarkable effect that adding
narrow-band filters does not reduce the achieved SNR. This makes it
possible to increase the SNR further by using spectral detectors that
can distinguish between different photon energies (within the limits
of the uncertainty principle) and effectively record many narrow
channels simultaneously. Or one can decrease the optical bandwidth so
much that photon count rates are sufficiently low not to be affected
by detector dead times.

In the NSII, the counting time $\dt$ was $\sim10$~nanosec.  Current
off-the-shelf instruments can achieve $\dt\simeq50$~picosec (not to
mention better quantum efficiency and broader wavelength response).
Clearly the time resolution is very good, but it is still orders of
magnitude longer than the coherence time. Both these inequalities are
important. The first inequality, or $\dt\gg\tcoh$, is what makes intensity
interferometry interesting in the first place.  In standard
interferometry, optical paths have to be kept under control to better
than $\lambda$, which is extremely demanding mechanically. For
intensity interferometry, path differences do not matter as long as
they are well below $c\dt$, corresponding to metres in the first
experiments. This not only relaxes the mechanical tolerances, it makes
the signal immune to atmospheric fluctuations.  Compared to other
techniques, intensity interferometry is possible with simpler
technology.  Put in another way, if we lack the precision to measure
the first-order correlation $G(x_1,x_2)$ directly, we can still get
information on it indirectly through the second-order correlation.
The second inequality, or $\dt\ll T$, is what makes intensity
interferometry usable with very low coincidence rates. The NSII could
operate at $r\,\tcoh \sim 10^{-5}$, because it had $10^8$ counting
times per second and hence could build up $\snr \sim 10^4 \,r\,\tcoh$
in a second.  Current technology could deliver at least another order
of magnitude better.

The above description, in terms of counting photons, is well suited to
optical astronomy.  In radio astronomy, on the other hand, a
description in terms of waves and intensities is standard. For
incoherent sources, the $\Epm$ fields are considered as (complex)
Gaussian random variables with autocorrelation functions determined by
the characteristics of the receiving system.  Each measurement
naturally consists of a finite sum of random amplitudes, so that the
fields and intensities vary with time even for sources of constant
luminosity. This `wave noise' or `self noise' defines the fluctuations
whose correlations are measured as the HBT effect. The wave noise also
adds to the photon shot noise that is important in the optical domain
of low photon rates, and this contribution actually dominates for high
$r\tcoh$.  We will return to this `Super-Poisson noise' \citep[see
  e.g.,][]{2006iai..book.....L} later.

Although the semi-classical picture is equivalent to the
quantum-optics picture on intensities versus photon
counts \citep{PhysRevLett.10.277}, a conceptual difficulty arises when
we consider the fields themselves.  In standard radio interferometry
(not HBT) the electric field is considered as a classical field which
can be measured directly.  But in quantum optics the electric field is
a non-Hermitian operator and hence not itself an observable.  How to
reconcile the radio-astronomy and quantum-optics pictures?  One
possible resolution is given in \cite{1969Burke}.  Here we suggest
another, which goes as follows.  Let there be a source field $\Spm$,
and let us superpose it on a known local field $\Lpm$.  The resulting
field
\begin{equation}
\Epm(x_1) = \Spm + \Lpm
\end{equation}
then gets its intensity measured:
\begin{equation} \label{Gsl}
G(x_1,x_1) = \big\langle \Sm \, \Sp \big\rangle +
             \big\langle \Lm \, \Lp \big\rangle +
             \big\langle \Sm \, \Lp \big\rangle +
             \big\langle \Lm \, \Sp \big\rangle \,\down.
\end{equation}
Here the first two terms on the right are just the intensities $|S|^2$
and $|L|^2$ of the two fields.  The last two terms give a beating
oscillation in the photon count rates, and it is from these
oscillations that the phase $\Spm$ is inferred.  Thus, even if the
source field cannot be measured directly, it can be inferred
indirectly.  To estimate the \snr, we note that the signal is in the
last two terms of \eqref{Gsl}, and the noise will be mainly the noise
in the second term.  Thus,
\begin{equation}
\snr \left(G(x_1,x_1)\right) \simeq |S|\,|L| \, / 
\mathop{\rm noise} \left( \big\langle \Lm \, \Lp \big\rangle \right) \,\down.
\end{equation}
If the local field is coherent, the denominator will follow Poisson
noise and hence be $\propto|L|$, making the \snr\ $\propto|S|$, just
as expected from the photon-counting picture.  If the local field is
not coherent, it will be subject to its own HBT fluctuations, or wave
noise.  We suggest the latter as a possible interpretation of the
well-known `receiver noise', which is the dominant noise in
radio astronomy.

\section{More than two detectors}

For chaotic sources, the higher order correlation functions are all
given in terms of the first-order correlations.  The result is presented
in equation~(10.27) of \cite{PhysRev.131.2766}, which, with a slight
modification of notation, reads
\begin{equation} \label{N-point}
G^{(n)}(x_1,\ldots,x_N,x_N,\ldots,x_1) =
\sum_{\cal P} \prod_{k=1}^N G(x_k,{\cal P}x_k) \,.
\end{equation}
Here ${\cal P}x_k$ denotes the $k^{\rm th}$ element of a permutation of
$\{x_1,\ldots,x_N\}$. The sum is over all permutations.  In each term,
the first argument always runs as $x_1,x_2,\dots$ whereas the second
argument runs as a permutation of that ordering.  As we remarked
earlier, in quantum optics, the $G$ are correlations between the field
operators, but the semi-classical approach of treating the fields as
classical and then interpreting intensities as photon probabilities is
valid for light \citep[cf.][]{PhysRevLett.10.277}.  Indeed,
equation \eqref{N-point} also appears in the classical theory of random
Gaussian variables, where it is known as Isserlis' theorem.

For three detectors, the formula \eqref{N-point} gives
\begin{equation}
\begin{aligned}
G^{(3)}(x_1,x_2,x_3,x_3,x_2,x_1) = {}
& G(x_1,x_1) \, G(x_2,x_2) \, G(x_3,x_3) +
  G(x_1,x_3) \, G(x_2,x_2) \, G(x_3,x_1) \, + \\
& G(x_1,x_2) \, G(x_2,x_3) \, G(x_3,x_1) +
  G(x_1,x_1) \, G(x_2,x_3) \, G(x_3,x_2) \, + \\
& G(x_1,x_3) \, G(x_2,x_1) \, G(x_3,x_2) +
  G(x_1,x_2) \, G(x_2,x_1) \, G(x_3,x_3) \,.
\end{aligned}
\end{equation}
In each term here, the first argument always runs as $x_1,x_2,x_3$
whereas the second argument runs as a permutation of that ordering.
One can rewrite this expression in another way, which is easier to
interpret, by introducing the normalized correlations
\begin{equation}
\begin{aligned}
g_{12}  &= \frac{G(x_1,x_2)G(x_2,x_1)}
                     {G(x_1,x_1) G(x_2,x_2)}  \,\up,  \\
g_{123} &= \frac{G(x_1,x_2) G(x_2,x_3) G(x_3,x_1)}
                     {G(x_1,x_1) G(x_2,x_2) G(x_3,x_3)} \,\up.
\end{aligned}
\end{equation}
The three-point coincidence rate is then
\begin{equation} \label{hbt3}
   1 + g_{12} + g_{13} + g_{23} + 2 \Re\,g_{123}
\end{equation}
times the chance coincidence rate.  Here the $g_{ij}$ terms are the
(normalized) power spectrum,\footnote{In terms of normalized
  visibilities $\gamma_{jk}$, we have $g_{12}=|\gamma_{12}|^2$ and
  $g_{123}=\gamma_{12}\gamma_{23}\gamma_{31}$.} while the last term is
the bispectrum, whose phase is well known in radio astronomy as the
closure phase.  Thus, whereas two-detector intensity interferometry
only measures amplitudes but no phases, multi-detector combinations
are actually sensitive to certain combinations of phases and provide
qualitatively new information.

If the detectors are close together, all the terms are equal
($g_{ij}=g_{123}=1$), resulting in a six-fold enhancement over the
chance coincidence rate.  This has been measured in lab experiments
\citep{PhysRevA.81.043831}.  At this point, one may get the idea, from
the $N!$ terms in \eqref{N-point}, that splitting up a single
collecting area into $N$ parts will increase the count rate.  But in
fact that will not happen.  Splitting reduces the count rate in each
detector by a factor of $N$, hence the $N$-point coincidence rate
would be $N!\,(r\,\tcoh/N)^N \propto \sqrt N {\rm e}^{-N} (r\,\tcoh)^N$, using
Stirling's approximation for large $N$.

Let us now estimate the \snr.  We have to be careful here, because the
combinatorial formulas define the coincidences over $\tcoh$.  We are
interested in coincidences over $\dt$, and we have seen before in the
case of two detectors that random and HBT coincidences scale
differently.  To take care of this, we have to understand that the
intensities that are being correlated are integrated over a duration
$\dt$ consisting of many intervals of $\tcoh$, and correlations exist
only for fields within the same short interval.  The $N$-point HBT
signal will therefore be
\begin{equation} \label{N-sig}
   g_{1.\,.\,.\,N} \times (r\,\tcoh)^N \, (\dt/\tcoh) \,.
\end{equation}
We should emphasize that \eqref{N-sig} is not the number of
coincidences, but the number that remains after subtracting off the
chance coincidence rate and all the lower-order HBT effects.
Meanwhile, the chance coincidence rate is
\begin{equation}
(r\,\dt)^N \,.
\end{equation}
Hence
\begin{equation} \label{N-snr}
\snr(N,\dt) \sim g_{1.\,.\,.\,N} \times
               (r\,\tcoh)^{N/2} \, (\tcoh/dt)^{N/2-1}
\end{equation}
For $N=2$ and $g_{12}\simeq1$ we recover the simple expression
\eqref{basic-snr} for standard HBT.

\section{Super-Poisson noise}

In deriving the expression \eqref{N-snr} for the SNR, we assumed that
the chance coincidence rate that determines the noise follows Poisson
statistics.  This is a good assumption for low count rates $r\,dt\ll
1$. But for very high photon rates, a new source of noise enters: it
is the well-known wave noise from radio-astronomy, and as mentioned in
the previous section, it can be considered as HBT correlation at a
single space point over different times. Even in the limit of
infinitely many photons, the received flux necessarily has to vary with
time, because it consists of a finite number of samples from a
stochastic process. These fluctuations that form the basis of the HBT
effect in the first place eventually also limit its achievable SNR.

As we mentioned above, the $N$-detector coincidence rate includes all
the lower-order HBT signals as well.  So, in principle, $N$-detector
coincidences could be used to extract the two-point HBT signal.  Is it
advantageous to do so, compared to standard HBT?  To answer this
question, let us consider the number of $N$-detector coincidences over
$\dt$
\begin{equation} \label{N-dots}
r^N \dt^N + r^N \dt^{N-1} \tcoh \sum_{j<k} g_{jk} + \dots
\end{equation}
The random coincidences are given by the first term, and two-point HBT
gives the next term. In the same way as before, we may derive
\begin{equation}
\snr(N\to2,\dt) \sim \Gamma \times (r\dt)^{N/2} \, \frac{\tcoh}{\dt} \,,
\end{equation}
where $\Gamma$ denotes the sum in equation \eqref{N-dots}.  Taking
many counting times in quadrature and assuming full correlation ($g_{jk}=1$), we have, analogous to equation \eqref{basic-snrT},
\begin{equation}
\snr(N\to2,T) \sim \frac {N(N-1)}{2}\, (r\dt)^{N/2} \, \frac{\tcoh}{\dt} \, \sqrt{\frac{T}{\dt}}  \; \up .
\end{equation}
If we keep the total collecting area constant but split it into $N$
detectors, so that the individual counting rate goes with $r/N$ we
have
\begin{equation} \label{snr-paradox}
\frac{\snr(N\to2,T)}{\snr(2,T)} =
(N-1)\,\left(\frac{r\dt}{N}\right)^{N/2-1} \,.
\end{equation}
In the photon-counting regime, $r\dt/N<1$, and hence splitting
detectors does not help.  The situation is different in the case of
$r\dt/N>1$, in which we may not be able to count individual photons,
but can still measure intensities. The \snr\ would then apparently
increase with $N$.  Formally, the optimal number of detectors for
$r\dt=1/10/100$ is $N=3/7/41$ with SNR benefits compared to two
detectors of $1.15/14.6/(1.42\times 10^9)$.  These numbers at first
look extremely promising.  But there is a paradox: equation
\eqref{snr-paradox} has \snr\ increasing with $\dt$ (that is, with
coarser time resolution), when $r\dt$ is large enough.  Clearly, this
estimate based on photon shot noise cannot represent the whole truth,
and resolving the paradox requires the inclusion of wave noise.

To see the effect of wave noise, let us return to the basic expression
\eqref{G37} of standard HBT, but now let $x_1,x_2$ be, not two
detector locations, but the same detector at different times.
Equation \eqref{G37} then means the mean-square intensity at a single
detector.  Let us take this mean square over a time interval
$\dt\gg\tcoh$.  The first term on the right of \eqref{G37} will
contribute $(r\dt)^2$.  The last term will contribute only if
$x_1,x_2$ happen to be closer than a coherence time.  There are
$\dt/\tcoh$ time slices over which that happens, and over each of
them, the last term contributes $(r\tcoh)^2$.  The full contribution
of the last term is hence $r^2\dt\,\tcoh$.  The expected photon count
at a single detector is thus not constant, but has a variance of
$r^2\dt\,\tcoh$.  Its square root is the single-detector wave noise
(photon shot noise comes on top of it) and is proportional to
intensity, not square-root intensity.  It is the single-detector wave
noise, also described as super-Poisson noise, and dominates if
$r\tcoh>1$.  For each sample of a Gaussian random process (like the
field of chaotic light) wave noise limits \snr\ to order unity.
Sampling the same field with larger collecting area does not
circumvent this fundamental limit. Even for strong sources, the final
\snr\ can thus never be higher than the square-root of the number of
samples.  The only chance to increase the \snr\ is observing for
longer or with more bandwidth.  This is a well-known (although often
neglected) effect in radio astronomy \cite[see
  e.g.,][]{1999ASPC..180..671R}.

Generalizing the $N$-point \snr\ formula \eqref{N-snr} to high photon
rates requires generalizing the above single-detector argument to $N$
detectors.  We leave a full calculation for a future paper, but the
basic conclusion, that \snr\ over one counting time $\dt$ cannot
exceed unity, remains valid for multiple detectors, no matter how
bright the source or how ridiculously large the light collectors.

\section{Fluxes and count rates}

From the previous section, we see that the number of photons
detectable in a coherence time, or $r\,\tcoh$, is central to the \snr.

In a passband around $\nu$, a source with flux density $F_\nu$ gives
\begin{equation}
   r\,\tcoh \simeq \frac{F_\nu}{h\nu} \, A,
\end{equation}
where $A$ is the collecting area. As mentioned before, the effect of
$\Delta\nu$ cancels between $r$ and $\tcoh$.  Rewriting the
expression, mixing wavelength and frequency, in the form
\begin{equation}
   r\,\tcoh  \simeq 0.05 \times
   \left(\frac{F_\nu}{\rm Jy}\right)
   \left(\frac\lambda{\rm m}\right)
   \left(\frac{A}{\rm m^2}\right)
\end{equation}
makes it easy to compare sources.

\begin{figure}
\begin{center}
\scalebox{0.6}{\includegraphics{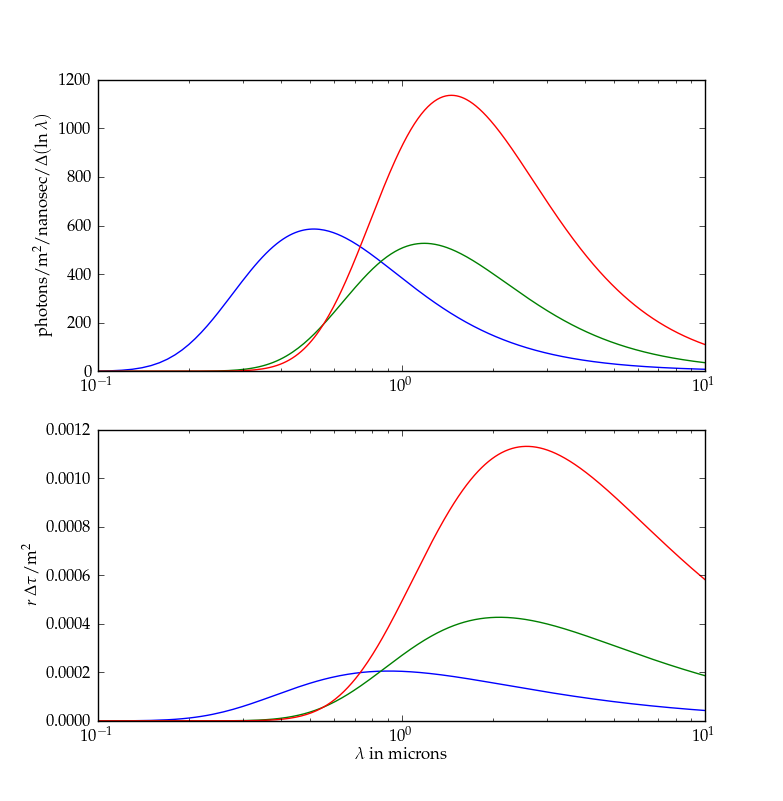}}
\end{center}
\caption{Photon flux (above) and photon flux per coherence time
  (below) for three blackbody sources, corresponding roughly to
  Sirius, Arcturus and Betelgeuse.  The lower panel leads to simple
  estimates of signal to noise.}
\label{fig:counts}
\end{figure}

Bright stars can have $F_\nu\sim10^{4}\rm\,Jy$, but
$\lambda\sim10^{-6}\rm\,m$.  Hence $r\,\tcoh\ll1$ per m$^2$.  More
detailed estimates are shown in Figure~\ref{fig:counts}.  On the other
hand, bright masers\footnote{For natural masers to show HBT
  correlation, it is essential that they are not single-mode systems
  like artificial masers, but rather like chaotic superpositions of
  laboratory masers. This fact also makes them spatially incoherent so
  that the usual Fourier relations for the correlations hold.} can
have $F_\nu\sim100\rm\,Jy$ at $\lambda\sim1\rm\,cm$, so a large dish
easily receives $r\,\tcoh>1$.  Now, intensity interferometry is not
needed to resolve masers, because at radio wavelengths it is much
easier to control the delays sufficiently accurately for standard
interferometry.  But high-order HBT using astrophysical masers would
be a novel quantum-optics experiment.

Blackbody sources lead to a rather elegant estimate of $r\,\tcoh$.
Consider a blackbody source at temperature $T$ and angular area
$\Omega$ on the sky.  We detect photons from it using a collector of
area $A$, in a passband $\Delta\lambda$.  For convenience, we will
work in terms of a logarithmic passband
\begin{equation}
\Delta (\ln\lambda) = \frac{\Delta\lambda}\lambda \;\up.
\end{equation}
We also write
\begin{equation}
z \equiv \frac{hc}{\lambda kT} \;\up.
\end{equation}
Recall the number density of a photon gas
\begin{equation}
    \frac{4\pi}{\lambda^3}
    \frac{\Delta(\ln \lambda)}{{\rm e}^z-1}  \;\up.
\end{equation}
This is for one polarization state; since we are making rough
estimates here, we will disregard the second polarization state.  To
get the photon flux, we multiply this by $c\Omega/(4\pi)$ or
$\lambda\nu\Omega/(4\pi)$.  The photon arrival rate in an area $A$ is
thus
\begin{equation}
r = \Omega A \, \frac\nu {\lambda^2}
    \frac{\Delta(\ln \lambda)}{{\rm e}^z-1}   \;\up .
\end{equation}
We can rewrite $\Delta\lambda$ in terms of the coherence time
(\ref{ctime}).  The number of photons received in a coherence time is
then
\begin{equation}
r\,\tcoh \sim \frac{\Omega A}{\lambda^2} \frac1{{\rm e}^z-1} \;\up.
\end{equation}
We can also write $r\,\tcoh$ in another way. Consider the baseline
needed to resolve the source. The square of the baseline is the area
$A_{\rm airy}\simeq \lambda^2/\Omega$ of an aperture whose
Airy disc corresponds to the size of the source. This is a natural
limit for the collecting area of detectors to just resolve the source
and, e.g., measure its size, because larger detectors would require
longer baselines and reduce the correlations $g$ so much that the
source would be `resolved out'.  With this definition, we have
\begin{equation} \label{airea}
r\,\tcoh \approx \frac A{A_{\rm airy}} \, \frac1{{\rm e}^z-1} \;\up.
\end{equation}
For very small $z$
\begin{equation}
r\,\tcoh \approx \frac A{A_{\rm airy}} \, \frac{kT}{hc} \lambda \,,
\quad \lambda \gg \frac{hc}{kT} \, \up.
\end{equation}
We see that for an HBT baseline adapted to the source size (that is,
$A/A_{\rm airy}$ fixed), the \snr\ only depends on $\lambda T$.  Thus,
by going far into the Rayleigh-Jeans tail, $r\tcoh$ can in principle
be made arbitrarily large (with the consequence of increasing wave
noise), but in the brightest part of the spectrum $r\tcoh\ll1$.

A simple physical interpretation for the expression~\eqref{airea} is
obtained by noting that it is $A/A_{\rm airy}$ times the phase-space
density of photons in blackbody radiation. As result of Liouville's
theorem, this phase-space density is conserved when the radiation
leaves the sources to travel towards the observer. When increasing the
distance, the total flux gets diluted, but at the same time the
apparent size of the source shrinks so that the momentum-space density
increases, which compensates for the former effect. $A_{\rm airy}$ is
a reciprocal measure for the apparent size of the source, and
$A/A_{\rm airy}$ by definition takes care of both effects. With
increasing distance, the light bucket has to grow to pick up more
photons, but can also form a smaller field of view and be susceptible
to a smaller part of momentum-space.

Figure~\ref{fig:counts} shows the count rate and $r\,\tcoh$ for three
blackbody sources.  These have $T=9940\rm\,K$ and angular diameter
$\theta=.007''$, which approximates Sirius, $T=4300\rm\,K$,
$\theta=.02''$, similar to Arcturus, and $T=3500\rm\,K$, and
$\theta=.04''$, similar to Betelgeuse.  \cite{1958RSPSA.248..222B}
measuring Sirius had a collecting area of about 2\,m$^2$, and quantum
efficiency 15\% at .4~micron.  From Figure~\ref{fig:counts} we would
predict $\snr\simeq3\times10^{-5}$ per counting cycle, less absorption
and other losses.  The reported value is $\snr=8.5$ in 345~minutes,
using 5--45~MHz, which corresponds to $\snr\simeq 1\times 10^{-5}$ per
counting cycle.

Equation~\eqref{airea} is not limited to bright stars, however.  As a
more exotic example of an approximately thermal source, consider the
central accreting region of M87 \citep[e.g.][]{2012Sci...338..355D}.
The innermost stable orbit lensed by the supermassive black hole has a
diameter of the order $50\rm\,\mu\,arcsec$. Resolving sources of this
size at optical wavelengths requires baseline lengths of several
kilometres. In order to achieve a similar $A/A_{\rm airy}$ as in the
NSII, collecting areas of several hectares would be needed.  Since
optical-quality mirrors are not required, such light buckets could be
built.  Whether such an observation is possible probably depends more
on how well the background light from the host galaxy can be excluded.

\section{Discussion}

We may expand the title of this paper to two questions.  First, is
the quantum-optical effect of three-point or higher-order HBT
measurable for any astronomical source?  If so, would it tell us new
things about the source?

Measuring three-point HBT for bright stars appears feasible.  Using
equation~\eqref{N-snr} for $N=2$ and $N=3$ scaled to an integration
time $T$,
\begin{equation} \label{snr-disc}
\begin{aligned}
\snr(T,N=2) &\sim g_{12} \times 
                  (r\tcoh) \sqrt{\frac{T}{\dt}} \, \up, \\
\snr(T,N=3) &\sim g_{123} \times
                  (r\tcoh)^{3/2} \frac{\sqrt{T\tcoh}}{\dt} \, \up.
\end{aligned}
\end{equation}
For visible light with a narrowband filter, the coherence time
$\tcoh\sim 10^{-12}\,$s.  Off-the-shelf photon counters can reach a
time resolution of $\dt\sim10^{-10}\,$s.  From
Figure~\ref{fig:counts}, a square metre of collecting area gives
$r\tcoh\sim10^{-4}$ from a bright star.  These numbers suggest
$\snr\sim1$ in an hour.  For quicker results one would want to
increase the collecting area --- from \eqref{N-snr}
we see that $T\propto A^{-N}$ for a given $\snr$.  As always in HBT, optical-path tolerances need only be $\ll c\dt$.

Three-point HBT would provide the three-point closure phase, or
bispectrum, of the source on the sky.  Is that worth having?  Since
two-point HBT gives only the power spectrum, and no phase information,
having the bispectrum is likely to be an important advantage for image
reconstruction of bright stars.

Going to four or more detectors, for bright stars or any other thermal
sources, would be difficult. The geometric estimate (\ref{airea}) of
the photon count per coherence time indicates that $r\,\tcoh \ll 1$
for any thermal source, except far in the Rayleigh-Jeans tail. But
$r\,\tcoh$ appears at progressively higher powers in the \snr.  For
non-thermal sources, the situation may be very different.  In
particular $r\,\tcoh>1$ appears achievable for bright masers.  Masers
can be imaged using standard radio-telescopes, so HBT may not provide
any new information on them.  Nonetheless, it would be an interesting
physics experiment to look for the combinatorial enhancement of the
HBT effect for large $N$.  The total number of photons would not
increase, of course, they would just get more and more bunched.

\section*{Acknowledgments}

The authors thank Dan D'Orazio, Richard Porcas and Tina Wentz for
discussions.

\bibliographystyle{mn2e}

\def\aap{A\&A}
\def\araa{ARA\&A}
\def\mnras{MNRAS}
\def\nat{Nature}
\def\nar{New Astr Rev}

\bibliography{heap}

\begin{thebibliography}{24}
\expandafter\ifx\csname natexlab\endcsname\relax\def\natexlab#1{#1}\fi

\bibitem[{{Baym}(1998)}]{1998AcPPB..29.1839B}
{Baym} G., 1998, Acta Physica Polonica B, 29, 1839

\bibitem[{{Borra}(2008)}]{2008MNRAS.389..364B}
{Borra} E.~F., 2008, \mnras, 389, 364

\bibitem[{{Burke}(1969)}]{1969Burke}
{Burke} B.~F., 1969, \nat, 223, 389

\bibitem[{{Doeleman} {et~al}\mbox{.}(2012){Doeleman}, {Fish}, {Schenck},
  {Beaudoin}, {Blundell}, {Bower}, {Broderick}, {Chamberlin}, {Freund},
  {Friberg}, {Gurwell}, {Ho}, {Honma}, {Inoue}, {Krichbaum}, {Lamb}, {Loeb},
  {Lonsdale}, {Marrone}, {Moran}, {Oyama}, {Plambeck}, {Primiani}, {Rogers},
  {Smythe}, {SooHoo}, {Strittmatter}, {Tilanus}, {Titus}, {Weintroub},
  {Wright}, {Young}, \& {Ziurys}}]{2012Sci...338..355D}
{Doeleman} S.~S. {et~al.}, 2012, Science, 338, 355

\bibitem[{{Dravins} {et~al}\mbox{.}(2012){Dravins}, {LeBohec}, {Jensen}, \&
  {Nu{\~n}ez}}]{2012NewAR..56..143D}
{Dravins} D., {LeBohec} S., {Jensen} H., {Nu{\~n}ez} P.~D., 2012, \nar, 56, 143

\bibitem[{{Foellmi}(2009)}]{2009A&A...507.1719F}
{Foellmi} C., 2009, \aap, 507, 1719

\bibitem[{Glauber(1963)}]{PhysRev.131.2766}
Glauber R.~J., 1963, Phys. Rev., 131, 2766

\bibitem[{Glauber(2006)}]{RevModPhys.78.1267}
Glauber R.~J., 2006, Rev. Mod. Phys., 78, 1267

\bibitem[{{Hanbury Brown}(1968)}]{1968ARA&A...6...13B}
{Hanbury Brown} R., 1968, \araa, 6, 13

\bibitem[{{Hanbury Brown} {et~al}\mbox{.}(1952){Hanbury Brown}, {Jennison}, \&
  {Das Gupta}}]{1952Natur.170.1061H}
{Hanbury Brown} R., {Jennison} R.~C., {Das Gupta} M.~K., 1952, \nat, 170, 1061

\bibitem[{{Hanbury Brown} \& {Twiss}(1956)}]{1956Natur.177.27H}
{Hanbury Brown} R., {Twiss} R.~Q., 1956, \nat, 177, 27

\bibitem[{{Hanbury Brown} \& {Twiss}(1958)}]{1958RSPSA.248..222B}
{Hanbury Brown} R., {Twiss} R.~Q., 1958, Royal Society of London Proceedings
  Series A, 248, 222

\bibitem[{{Horch} \& {Camarata}(2012)}]{2012SPIE.8445E..2LH}
{Horch} E.~P., {Camarata} M.~A., 2012, in Society of Photo-Optical
  Instrumentation Engineers (SPIE) Conference Series, Vol. 8445, Society of
  Photo-Optical Instrumentation Engineers (SPIE) Conference Series

\bibitem[{{Jain} \& {Ralston}(2008)}]{2008A&A...484..887J}
{Jain} P., {Ralston} J.~P., 2008, \aap, 484, 887

\bibitem[{{Kiesel} {et~al}\mbox{.}(2002){Kiesel}, {Renz}, \&
  {Hasselbach}}]{2002Natur.418..392K}
{Kiesel} H., {Renz} A., {Hasselbach} F., 2002, \nat, 418, 392

\bibitem[{{Labeyrie} {et~al}\mbox{.}(2006){Labeyrie}, {Lipson}, \&
  {Nisenson}}]{2006iai..book.....L}
{Labeyrie} A., {Lipson} S., {Nisenson} P., 2006, {An Introduction to
  Astronomical Interferometry}. Cambridge University Press

\bibitem[{{Nu{\~n}ez} {et~al}\mbox{.}(2012){Nu{\~n}ez}, {Holmes}, {Kieda}, \&
  {Lebohec}}]{2012MNRAS.419..172N}
{Nu{\~n}ez} P.~D., {Holmes} R., {Kieda} D., {Lebohec} S., 2012, \mnras, 419,
  172

\bibitem[{{Ofir} \& {Ribak}(2006{\natexlab{a}})}]{2006MNRAS.368.1646O}
{Ofir} A., {Ribak} E.~N., 2006{\natexlab{a}}, \mnras, 368, 1646

\bibitem[{{Ofir} \& {Ribak}(2006{\natexlab{b}})}]{2006MNRAS.368.1652O}
{Ofir} A., {Ribak} E.~N., 2006{\natexlab{b}}, \mnras, 368, 1652

\bibitem[{{Radhakrishnan}(1999)}]{1999ASPC..180..671R}
{Radhakrishnan} V., 1999, in Astronomical Society of the Pacific Conference
  Series, Vol. 180, Synthesis Imaging in Radio Astronomy II, {Taylor} G.~B.,
  {Carilli} C.~L., {Perley} R.~A., eds., p. 671

\bibitem[{Sato {et~al}\mbox{.}(1978)Sato, Wadaka, Yamamoto, \& Ishii}]{Sato:78}
Sato T., Wadaka S., Yamamoto J., Ishii J., 1978, Appl. Opt., 17, 2047

\bibitem[{Sim {et~al}\mbox{.}(2012)Sim, Cheng, Bessarab, Jones, \&
  Krivitsky}]{PhysRevLett.109.113601}
Sim N., Cheng M.~F., Bessarab D., Jones C.~M., Krivitsky L.~A., 2012, Phys.
  Rev. Lett., 109, 113601

\bibitem[{Sudarshan(1963)}]{PhysRevLett.10.277}
Sudarshan E. C.~G., 1963, Phys. Rev. Lett., 10, 277

\bibitem[{Zhou {et~al}\mbox{.}(2010)Zhou, Simon, Liu, \&
  Shih}]{PhysRevA.81.043831}
Zhou Y., Simon J., Liu J., Shih Y., 2010, Phys. Rev. A, 81, 043831

\end{thebibliography}

\end{document}